\documentclass[superscriptaddress,twocolumn,pra,aps,showpacs,]{revtex4-1}
\usepackage{epsfig}
\usepackage{bm,hyperref}
\usepackage{amssymb}
\usepackage{mathrsfs}
\usepackage{amsmath}
\usepackage{CJK}

\begin{document}
\title{Optically trapped quasi-two-dimensional Bose gases in random environment: quantum fluctuations and superfluid density}
\author{Kezhao Zhou}
\affiliation{Shenyang National Laboratory for Materials Science,
Institute of Metal Research, and International Centre for Materials
Physics, Chinese Academy of Sciences, 72 Wenhua Road, Shenyang
110016, People's Republic of China}
\author{Ying Hu}
\affiliation{Department of Physics, Centre for Nonlinear Studies,
and The Beijing-Hong Kong-Singapore Joint Centre for Nonlinear and
Complex Systems (Hong Kong), Hong Kong Baptist University, Kowloon
Tong, Hong Kong, China}
\author{Zhaoxin Liang}
\email{zhxliang@gmail.com}
\affiliation{Shenyang National Laboratory
for Materials Science, Institute of Metal Research, and
International Centre for Materials Physics, Chinese Academy of
Sciences, 72 Wenhua Road, Shenyang 110016, People's Republic of
China}
\author{Zhidong Zhang}
\affiliation{Shenyang National Laboratory for Materials Science,
Institute of Metal Research, and International Centre for Materials
Physics, Chinese Academy of Sciences, 72 Wenhua Road, Shenyang
110016, People's Republic of China}
\date{\today}

\begin{abstract}
We investigate a dilute Bose gas confined in a tight one-dimensional
(1D) optical lattice plus a superimposed random potential at zero
temperature. Accordingly, the ground state energy, quantum depletion
and superfluid density are calculated. The presence of the lattice
introduces a crossover to the quasi-2D regime, where we analyze
asymptotically the 2D behavior of the system, particularly the effects
of disorder. We thereby offer an analytical expression for the
ground state energy of a purely 2D Bose gas in a random potential.
The obtained disorder-induced normal fluid density $n_n$ and quantum
depletion $n_d$ both exhibit a characteristic
$1/\ln\left(1/n_{2D}a_{2D}^{2}\right)$ dependence. Their ratio
$n_n/n_d$ increases to $2$ compared to the familiar $4/3$ in
lattice-free 3D geometry, signifying a more pronounced contrast
between superfluidity and Bose-Einstein condensation in low
dimensions. Conditions for possible experimental realization of our
scenario are also proposed.

\end{abstract}

 \maketitle

\section{Introduction}

The effect of dimensionality of a bosonic system on the presence and
nature of the Bose-Einstein condensation (BEC) as well as on the
superfluid phase transition has received long-standing interests
both experimentally and theoretically \cite{Petrev,Posa}. The
physics at low dimensions exhibits fundamental difference from that
in three-dimension (3D). In particular, the strong long-range phase
fluctuations typical of low-dimensional bosonic systems usually
inhibit the formation of long-range order, which on the other hand
characterizes the 3D BEC and corresponding phase transition at low
temperature \cite{Pitbook}.

Earlier work on low-dimensional bosonic systems \cite{Posa} have
culminated in, particularly in uniform 2D case, two important
theoretical discoveries. The first is that in 2D a true condensate
can only occur at $T=0$ and its absence at finite temperature
follows from Bogoliubov $k^{-2}$ \cite{Bigoliu} or
Hohenberg-Mermin-Wagner (BHMW) theorem \cite{Mermin,Hohenberg}. On
the other hand, a superfluid phase transition has been proven to
exist at sufficiently low temperature in 2D \cite{Kane,Berez}.
However, according to Kosterlitz and Thouless (KT) \cite{KI}, such
transition is associated with the unbinding of vortex pairs or
quasi-long-range order, in contrast to the 3D phase transition that
features long-range order parameter. Below the KT transition
temperature, a 2D Bose gas (liquid) is characterized by the presence
of a ``quasicondensate" \cite{Popov,Mora}.

The remarkable experimental progress with ultracold atomic gases,
especially in cooling and confining of cold atomic gases in traps
with controllable geometry and dimension, have significantly
stimulated new interests in low-dimensional systems \cite{Bloch}.
Tight confinement in one or two directions considerably affects the
properties of Bose gases such as collisions and phase fluctuations
\cite{Petrov}, introducing a crossover to the quasi-low-dimensional
regime. As such, quasi-2D quantum degenerate Bose gases have been
experimentally produced both in single ``pancake" traps and at the
nodes of 1D optical lattice \cite{Exp2D}.

However, these marginal 2D Bose gases are qualitatively different
from corresponding infinite ones. Along this line, Petrov {\it et
al.} \cite{Petrov} have pointed out that the presence of trapping
potential suppresses long-range thermal fluctuations and that in a
quasi-2D system a true condensate can exist within a wide parameter
range. This theoretical prediction has been echoed by Fischer {\it
et al.} \cite{Fischer} who obtained in a marginal 2D case a
model-independent geometrical equivalence of the BHMW theorem.

Compared with harmonically trapped systems, optically trapped Bose gases
allow more experimental controllability with tunable inter-atomic
interactions, tunneling amplitudes between adjacent sites, atom
filling fractions and lattice dimensionality \cite{Bloch}, thereby
presenting a more useful testing ground for theoretical ideas in
studying low-dimensional systems in novel conditions. On the other
hand, disorder has been observed to cause dramatic influence on a
BEC and has attracted huge interests recently \cite{Disorderrev1,Disorderrev2}. In view of the
availability to control a 1D optical lattice and external
randomness, therefore, one especially appealing direction of
investigation consists in studying the effect of external randomness
on a Bose gas trapped in a 1D optical lattice.

In this paper, we investigate the ground state properties and
superfluidity of a 1D-optical-lattice trapped Bose gas in a random
environment at $T=0$. Capitalizing on the characteristic
lattice-induced 3D to quasi-2D dimensional crossover, we analyze
effects of disorder in the asymptotic 2D regime. The present work is
composed of two parts. In the first part, we calculate the ground
state energy and quantum depletion for the model system using the
path integral approach within the Bogoliubov approximation.
Discussion on the dimensional crossover property in a random
potential is presented. In particular, our results in the quasi-2D
regime with varnishing disorder are in good agreement with that of a
homogeneous 2D Bose gas at $T=0$ \cite{Posa,Schick}. We suggest,
therefore, that our result gives the analytical expression for the
ground state energy of a uniform 2D dilute Bose gas in the presence
of weak disorder. In the second part, we calculate the disorder and
lattice induced normal fluid density $n_n$ at $T=0$. Our results in
the anisotropic 3D regime reproduce the well known ratio
$n_n/n_d=4/3$ \cite{Huang,Hu} with $n_d$ being the quantum depletion
due to disorder. Whereas, in the quasi-2D regime, $n_n$ exhibits a
$1/\ln \left(1/n_{2D} a_{2D}^{2}\right)$ dependence unique of a 2D
system and the ratio becomes asymptotically $n_n/n_d=2$, indicating
a more pronounced contrast between superfluidity and BEC in low
dimensions.

The outline of the paper is as follows. In Sec. II,  we introduce
the grand canonical partition function for a dilute Bose gas in the
presence of a 1D optical lattice and weak disorder at $T=0$.
Accordingly, the analytical expressions for the ground state energy
and quantum depletion are derived. Sec. III presents a detailed
discussion on the dimensional crossover in the ground state
properties induced by a 1D optical lattice. Effects of disorder in
the crossover regimes are analyzed. In Sec. IV, we calculate the
superfluid density and study its behavior respectively in the 3D and
quasi-2D regime. Finally, we summarize our results in Sec. V and
propose possible experimental scenarios.

\section{Bose gases in the presence of a 1D optical lattice and weak disorder}

\subsection{Path integral approach}

Our starting point is the grand-canonical partition function of a 3D
weakly interacting dilute Bose gas \cite{Popovbook} in the presence of a 1D optical
lattice and weak disorder
\begin{equation}\label{PF}
Z=\int D\left[\psi^{*},\psi \right] e^{-\frac{S\left[\psi^{*},\psi\right]}{\hbar}},
\end{equation}
where the action functional $S\left[\psi^{*},\psi\right]$ reads
\begin{eqnarray}
\!\!S\left[\psi^{*},\psi\right]&=&\int_{0}^{\hbar \beta }d\tau \int
d\mathbf{r}\psi^{*} (\mathbf{r},\tau )\Bigg[\hbar \frac{\partial
}{\partial \tau }-
\frac{\hbar ^{2}\nabla^{2}}{2m}-\mu  \nonumber \\
&+&V_{opt}(\mathbf{r})\!+\!V_{ran}(\mathbf{r})\!+\!\frac{g_{e}}{2}|\psi(
\mathbf{r},\tau )|^2\Bigg]\psi (\mathbf{r},\tau ).  \label{Action}
\end{eqnarray}
In Eqs. (\ref{PF}) and (\ref{Action}),
$\left[\psi^{*}\left(\mathbf{r}
,\tau\right),\psi\left(\mathbf{r},\tau\right)\right]$ collectively
denote the complex functions of space and imaginary time $\tau$,
$\beta=1/k_BT$ with $k_B$ being the Boltzmann constant and $T$ being
the temperature, $\mu$ is the chemical potential and $g_{e}$ is the
effective two-body coupling constant in the presence of a 1D optical
lattice. The $V_{opt}(\mathbf{r})$ and $V_{ran}(\mathbf{r})$
respectively represent the 1D optical lattice and external random
potential.

The optical potential $V_{opt}(\mathbf{r})$ in Eq. (\ref{Action}) is
given by
\begin{equation}\label{Vopt}
V_{opt}(\mathbf{r})=s\times E_{R}\sin ^{2}(q_{B}z),
\end{equation}
where $s$ is a dimensionless factor labeled by the intensity of
laser beam and $E_{R}=\hbar ^{2}q_{B}^{2}/2m$ is the recoil energy
with $\hbar q_{B}$ being the Bragg momentum. The lattice period is
fixed by $q_{B}=\pi /d$ with $d$ being the lattice spacing. Atoms
are unconfined in the $x-y$ plane.

Disorder $V_{ran}(\mathbf{r})$ in Eq. (\ref{Action}) is produced by
the random potential associated with quenched impurities
\cite{Huang,Hu,Astra}
\begin{equation}  \label{Ran}
V_{dis}(\mathbf{r})=\sum_{i=1}^{N_{imp}}v\left(|\mathbf{r-r_{i}}|\right),
\end{equation}
with $v(\mathbf{r})$ describing the two-body interaction between
bosons and impurities, $\mathbf{r}_i$ being the randomly distributed
positions of impurities and $N_{imp}$ counting the number of
$\mathbf{r}_i$. Here, we restrict ourselves to the conditions of a
dilute BEC system in the presence of a very small concentration of
disorder. Thereby, $v(\mathbf{r })$ can be approximated by an
effective pseudo-potential in the form $v(\mathbf{r
})=g_{imp}\delta(\mathbf{r})$ \cite{Huang}, with $g_{imp}$ being the
effective coupling constant of an impurity-boson pair confined in a
1D optical lattice.

It's important to mention that the tight confinement in the
direction of optical lattice considerably influences the value of
effective coupling constant \cite{Petrov,Fedichev} in Eq.
({\ref{Action}}). Particularly, in the presence of optical lattice,
$g_e$ generally exhibits dependence on the density and lattice
parameter \cite{Wouters}, in marked contrast to a free 3D Bose gas
where $g_{3D}=4\pi\hbar^2a_{3D}/m$ with $a_{3D}$ being the 3D
scattering length. For formulation clarity, however, below we shall
use $g_e$ and $g_{imp}$ for notational convenience while leaving
aside their specific expressions in order to obtain general
expressions for the ground state energy and quantum depletion.
Analysis of the lattice-renormalized effective coupling constant
will be given in Section IV.

\subsection{Beyond-mean-field ground state energy and quantum depletion}

In what follows, we focus on the situation where the optical lattice
is strong enough to create many separated wells that give rise to an
array of condensates; while full coherence is still assured by the
quantum tunneling. By this assumption, one can refer to $n_0$ as the
condensate density and neglect the Mott insulator phase transition.
We also suppose disorder is sufficiently weak. Under these
conditions, one is able to investigate the ground state properties
of the model system using Bogoliubov's theory \cite{Pitbook}.

We shall restrict ourselves to the case where $s$ is relatively
large that the interwell barriers are significantly higher than the
chemical potential $\mu$ \cite{Orso}. We thereby only consider the lowest Bloch
band where the condensate, in the tight-binding approximation, can
be written in terms of Wannier functions as
$\phi_{k_z}(z)=\sum_{l}e^{ilk_z}w(z-ld)$ where $w(z)=\exp(-z^2/{
2{\sigma^2}})/\pi^{1/4}\sigma^{1/2}$ with $d/\sigma \simeq\pi
s^{1/4} \exp(-1/4\sqrt{s})$. Expanding the bosonic field variables
in Eq. ({\ref{Action}) by the expression
$\psi\left(\mathbf{r},\tau\right)=\sum_{\mathbf{k},n}\psi _{
{\bf k},n}\phi_{k_z}(z)e^{-i(k_xx+k_yy)}e^{i\omega _{n}\tau}$ with $
\omega _{n}=2\pi n/\hbar \beta$ being the bosonic Matsubara
frequencies where $n$ are integers, the action Eq. ({\ref{Action}})
takes the form
\begin{eqnarray}  \label{Action1}
\frac{S\left[\psi^{*},\psi \right]}{\hbar \beta V}
&=&\sum_{\mathbf{k} ,n}\psi^{*}_{\mathbf{k},n} \left[-i\hbar\omega
_{n}+\varepsilon _{\mathbf{k}
}^0-\mu \right] \psi _{\mathbf{k},n}  \nonumber \\
&+&\frac{\tilde{g}_{e}}{2}\sum_{\substack{
\mathbf{k},\mathbf{k}^{\prime }, \mathbf{q}\nonumber \\ n,n^{\prime
},m}}\psi^{*}_{\mathbf{k}+\mathbf{q}
,n+m}\psi^{*}_{\mathbf{k}^{\prime }-\mathbf{q},n^{\prime}-m}\psi
_{\mathbf{k}
^{\prime },n^{\prime }}\psi _{\mathbf{k},n}  \nonumber \\
&+&\sum_{\mathbf{k},\mathbf{k}^{\prime
},n}V_{\mathbf{k}-\mathbf{k}^{\prime
}}\psi^{*}_{\mathbf{k},n}\psi_{\mathbf{k}^{\prime },n}.
\end{eqnarray}
Here, $\varepsilon _{{\bf k}}^0=(\hbar
^{2}/2m)(k_x^2+k_y^2)-2t[1-\cos (k_{z}d)]$, with $t$ being the
tunneling rate between neighboring wells, is the energy dispersion
of the noninteracting model, $V$ is the volume of the system and
$\tilde{g}_e$ is the lattice renormalized coupling constant given by
\begin{equation}\label{Ge}
\tilde{g}_e=g_e\left[d\int_{-d/2}^{d/2}w^4(z)dz\right]=g_e\frac{d}{\sqrt{2\pi}\sigma}.
\end{equation}
In Eq. (\ref{Action1}), the $V_{k}$ is the Fourier transform of
$\tilde{V} _{ran}(\mathbf{r})=\sum_i
\tilde{g}_{imp}\delta(\mathbf{r}-\mathbf{r}_i)$ with
$\tilde{g}_{imp}=g_{imp}d/\sqrt{2\pi}\sigma$ being the
lattice-renormalized impurity-boson coupling constant, i.e.
$V_{\mathbf{k}}=(1/V)\int e^{i\mathbf{k}\cdot \mathbf{r}}\tilde{V}
_{ran}(\mathbf{r})d\mathbf{r}$. For simplicity, the external
randomness is assumed to be uniformly distributed with density
$n_{imp}=N_{imp}/V$ and Gaussian correlated \cite{Falco}. Hence the
two basic statistical properties of the disorder are the average
value $ \langle V_0\rangle=\tilde{g}_{imp}n_{imp}$ and the
correlation function $ \langle
V_{-\mathbf{k}}V_{\mathbf{k}}\rangle=\tilde{g}_{imp}^2n_{imp}/V$.
Here, the notation $\langle..\rangle$ stands for the ensemble
average over all possible realization of disorder configurations.

By applying the Bogoliubov theory to the action ({\ref{Action1}})
and proceeding in the standard fashion \cite{Pitbook}, one obtains
the zero-temperature thermodynamic function $\Omega=E_g-V\mu n_0$
with the ground state energy $E_g$ reading
\begin{eqnarray}
\frac{E_g}{V}&&=\frac{1}{2}\tilde{g}_en_0^2-\frac{1}{2V}\sum_{{\bf k}\neq 0}\left(\varepsilon_{\mathbf{k}}^0+\tilde{g}_en_0-E_{\mathbf{k}}\right)\nonumber\\
&+&n_0\left[n_{imp}\tilde{g}_{imp}-\frac{n_{imp}\tilde{g}_{imp}^2}{V}
\sum_{{\bf k}\neq 0}\frac{\varepsilon_{\mathbf{k}}^0}{E_{\mathbf{k}}^2}\right].\label{Eg0}
\end{eqnarray}
Here, $E_{\mathbf{k}}=\sqrt{ (\varepsilon _{\mathbf{k}}^0-\mu
+2\tilde{g}_{e}n_{0})^{2}-\tilde{g}_{e} ^{2}n_{0}^{2}}$ is the
energy spectrum of the elementary excitations and $n_0$ is the
condensate density. In conformity with the general theory, we set $\mu=\tilde{g}_{e}n_0$
to ensure a gapless quasiparticle spectrum \cite{hugenholtz}.

In the continuum limit, the sum in Eq. ({\ref{Eg0}}) is replaced
with integrals. To avoid the large-$k$ divergence in the integration
over $k_x$ and $k_y$, however, one must introduce a renormalization
of the coupling constant by replacing $\tilde{g}_{e}\rightarrow
\tilde{g}_{e}-\left(\tilde{g}_e^2/V\right)\sum_{\mathbf{k}\neq0}\left(1/2\varepsilon_{\mathbf{k}}^0\right)$
and $ \tilde{g}_{imp}\rightarrow
\tilde{g}_{imp}-\left(\tilde{g}_{imp}^{2}/V\right)\sum_{{\bf k} \neq
0}\left(1/2\varepsilon_{\mathbf{k}}^0\right)$ in Eq. ({\ref{Eg0}}). Upon this
replacement, one obtains after integration
\begin{eqnarray}
\frac{E_g}{V}=\frac{1}{2}\tilde{g}_en_0^2\Bigg\{(&1&+\gamma)+\frac{m\tilde{g}_e}{2\pi^2\hbar^2d}F\left(\frac{2t}{\tilde{g}_en_0}\right)
\nonumber\\
&+&\frac{m\tilde{R}\tilde{g}_e}{2\pi\hbar^2d}\text{arccoth}\left[\left(\frac{2t}{\tilde{g}_en_0}+1\right)^{\frac{1}{2}}\right]\Bigg\},\label{Eg}
\end{eqnarray}
where the two parameters $\gamma=2\kappa\tilde{g}_{imp}/\tilde{g}_e$
with $\kappa=n_{imp}/n_0$ and
\begin{equation}\label{R}
\tilde{R}=\frac{n_{imp}}{n_0}\frac{4\tilde{g}_{imp}^2}{\tilde{g}_e^2}
\end{equation}
characterize the strength of disorder in a 1D optical lattice. In
Eq. ({\ref{Eg}}), the function $F(x)$ with the variable
$x=2t/\left(\tilde{g}_en_0\right)$ is defined as
\begin{eqnarray}
F(x)
&=&\frac{(x+1)}{2}\left[\left(3x+1\right)\arctan\left(\frac{1}{\sqrt{x}}\right)-3\sqrt{x}
\right]  \nonumber \\
&-&\frac{\pi}{2}\ln \left[\frac{x}{2x+1+2\sqrt{x\left(x+1\right)}}\right] \nonumber\\
&-&\pi\text{arcsinh}\left(\sqrt{x}\right)+2\int_{0}^{\sqrt{x}}\frac{\tan
^{-1}(z)}{z}dz.\label{ff}
\end{eqnarray}

\begin{figure}
\includegraphics[width=\columnwidth,height=50mm]{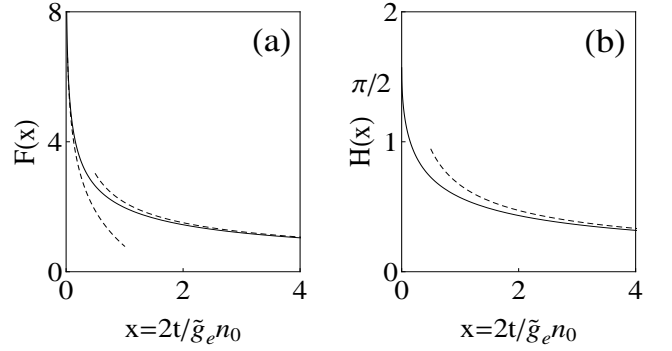}
\caption{\label{fh} (a): Scaling function F(x) in Eq. (\ref{ff}) (solid line) and its
asymptotic behavior (dashed line). (b): Scaling function H(x) in Eq. (\ref{hh}) (solid
line) and its asymptotic behavior (dashed line).}\label{ffhh}
\end{figure}

The integration in Eq. (\ref{ff}) can be easily performed
numerically and the result is shown in Fig. \ref{ffhh}(a). In the
ground state energy Eq. ({\ref{Eg}}), the first two terms give the
mean-field contribution modified by an optical lattice and disorder;
whereas, the last two terms represent beyond-mean-field corrections,
as a consequence of quantum fluctuations respectively induced by
interatomic interaction and external randomness.

Quantum depletion $\left(\Delta N=N-N_0\right)$ refers to the
average number of atoms with nonzero momentum \cite{Pitbook} which
can be calculated within the Bogliubov's theory as
\begin{eqnarray}
\Delta N\!=\!\sum_{\mathbf{k}\neq 0}\Bigg[\frac{\varepsilon^0
_{\mathbf{k}}+\tilde{g}_{e}n_0\!-\!E_{\mathbf{k}}}{2E_{\mathbf{k}}}+n_0
n_{imp}\tilde{g}_{imp}^{2} \frac{\left(\varepsilon
_{\mathbf{k}}^0\right)^{2}}{E_{\mathbf{k}}^{4}}\Bigg]. \label{Np}
\end{eqnarray}
By replacing the sum with the integral in the continuum limit, one
obtains
\begin{equation}
\frac{\Delta N}{N}=\frac{m\tilde{g}_e}{2\pi^2 \hbar^2d}\left[H\left(\frac{2t}{\tilde{g}_{e}n_0
}\right)+\frac{\pi\tilde{R}}{8}\left(1+\frac{2t}{\tilde{g}_en_0}\right)^{-\frac{1}{2}}\right], \label{Depletion}
\end{equation}
where the function $H(x)$ with $x=2t/\left(\tilde{g}_en_0\right)$ is defined as
\begin{equation}\label{hh}
H(x)=\left(x+1\right)\arctan\left(\frac{1}{\sqrt{x}}\right)-\sqrt{x}.
\end{equation}

\section{Dimensional crossover from 3D to quasi-2D and 2D regimes}

At low energies, the physical properties of a dilute Bose gas can be
expressed in terms of the two-body scattering amplitude
\cite{Lifshitz}. It has been well established that a tight
confinement along one or two directions will considerably affect the
scattering properties of atoms, particularly, introducing a
dimensional crossover from anisotropic 3D to low-dimensional regimes
\cite{Petrov,Wouters}.

The two-body scattering problem in the presence of a 1D optical
lattice has been analytically investigated in Ref. \cite{Wouters}.
For sufficiently deep lattices and chemical potential $\mu$, which
is small compared to the interband gap, two distinct regimes can be
identified: (i) for $\mu\ll 4t$, where the wavefunction spreads over
many lattice sites, the system retains an anisotropic 3D behavior.
In this limit, Eq. (\ref{Ge}) takes the limiting form
$\tilde{g}_e= \tilde{g}_{3D}$ with
\begin{equation}  \label{g3D}
\tilde{g}_{3D} =\frac{4\pi\hbar^2\tilde{a}_{3D}}{m},
\end{equation}
with $\tilde{a}_{3D}=a_{3D}d/(\sqrt{2\pi}\sigma)$ being the
lattice-renormalized $s$-wave scattering length; (ii) for $\mu\gg
4t$, the tunneling between wells is negligible, and the two
interacting bosons are in the ground state of an effective harmonic
potential characterized by frequency $\omega_0=\hbar/m\sigma^2$ and
harmonic oscillator length $\sigma$. In this limit, the system
undergoes a crossover to the quasi-2D regime where the coupling
constant is reduced to that in a tight confined harmonic trap
$\tilde{g}_e=g_{h}d$ \cite{Wouters,Petrov,pilati} where
\begin{equation}\label{gh}
g_{h}=
\frac{2\sqrt{2\pi}\hbar^2}{m}\frac{1}{a_{2D}/a_{3D}+(1/\sqrt{2\pi})\ln{\left[1/n_{2D}a_{2D}^2\right]}},
\end{equation}
with the surface density $n_{2D}=n_0d$ and the effective 2D
scattering length $a_{2D}=\sqrt{\hbar/m\omega_0}=\sigma$
\cite{Petrov}. With decreasing $\sigma$, the 2D features in the
scattering of two atoms become pronounced \cite{Petrov}. And in the
limit $\sigma\ll a$, Eq. ({\ref{gh}}) becomes independent of the
value of $a_{3D}$ and a regime of purely 2D scattering is achieved
with Eq. (\ref{gh}) reducing to the coupling constant of a purely 2D
Bose gas $g_h\rightarrow g_{2D}$ where
\begin{equation}  \label{g2D}
g_{2D}=\frac{4\pi \hbar ^{2}}{m}\frac{1}{\ln (1/n_{2D}a_{2D}^{2})}.
\end{equation}
Here the logarithmic dependence on the gas parameter
$n_{2D}a_{2D}^2$ is unique of the 2D geometry.

Taking into account of the dimensional crossover in the effective
coupling constant, below we focusing on analyzing the behavior of
the ground state energy in Eq. (\ref{Eg}) and quantum depletion in
Eq. ({\ref{Depletion}}), respectively in the anisotropic 3D and 2D
geometry. In the limit $2t/n_0\tilde{g}_e\gg 1$, corresponding to
the anisotropic 3D regime, we find $F(x)\simeq 32/15 \sqrt{x}$, as
is shown in Fig. \ref{ffhh}(a) with the dashed curve. Substitutions
of this limiting value in Eq. (\ref{Eg}) together with Eq.
({\ref{g3D}}), yield the ground state energy of an effectively free
3D Bose gas composed of bosons with effective mass
$m^{*}=\hbar^2/2td^2$ and coupling constant $\tilde{g}_{3D}$
\cite{Kramer}
\begin{eqnarray}
\frac{E_g}{V}=\frac{1}{2}\tilde{g}_{3D}n_0^2\Bigg[\Big(
1\!&+&\!\!\kappa
\frac{\tilde{b}_{3D}}{\tilde{a}_{3D}}\Big)\!+\!\!\frac{128}{
15}\!\!\sqrt{\frac{m^{*}}{m
}}\left(\frac{n_0\tilde{a}_{3D}^3}{\pi}\right)^{1/2}  \notag \\
&+&4\pi\tilde{R}_{3D}\!\!\sqrt{\frac{m^{*}}{m
}}\left(\frac{n_0\tilde{a}_{3D}^3}{\pi}\right)^{1/2}\Bigg].
\label{E3D}
\end{eqnarray}
In Eq. (\ref{E3D}), the two characteristic parameters of disorder in
Eq. (\ref{Eg}) respectively take their 3D value, i.e.
$\gamma=\kappa\tilde{b}_{3D}/\tilde{a}_{3D}$ and and
$\tilde{R}_{3D}=\kappa\tilde{b}_{3D}^2/\tilde{a}_{3D}^2$, showing
the 3D feature of the interaction between the impurity-boson pair.
The first term in Eq. (\ref{E3D}) represents the mean-field ground
state energy; whereas the remaining terms exhibit the familiar
dependence on the effective 3D gas parameter $\sqrt{n_0
\tilde{a}_{3D}^{3}}$, thereby consisting of the generalized LHY
correction \cite{Lee} to the presence of a 1D optical lattice and
weak disorder. Eq. ({\ref{E3D}}) bears formal resemblance with the
corresponding result in Ref. \cite{Hu} which deals with a 2D optical
lattice system, in consistent with the effective mass theory in the
3D limit where the lattice system is effectively treated as a free
gas with effective mass and coupling constant. The main difference
is related to the value of renormalized coupling constant
$\tilde{g}_{3D}$ where the renormalization factor for different
lattice dimensions \cite{Kramer}.

In the opposite 2D regime where $2t/\tilde{g}_{e}n_{0}\ll 1$ and
$\sigma\ll a$, $F(x)$ exactly approaches a limit $F(x)=\pi /4-\pi
/2\log x$ with $\log x\simeq \ln (mt/n_{2D}2\pi \hbar ^{2})+\ln
\left[\ln \left(1/n_{2D}a_{2D}^{2}\right)\right]$, as shown in Fig.
\ref{ffhh}(a) with the dashed line. In this limit, the Bloch
dispersion can be neglected and the scattering problem reduces to 2D
with the coupling constant Eq. ({\ref{g2D}}). In such conditions,
Eq. ({\ref{Eg}}) yields the ground state energy of a 2D Bose gas in
the presence of disorder
\begin{eqnarray}\label{eg2d}
\frac{E_{g2D}}{L^2} &\simeq
&\frac{1}{2}g_{2D}n_{2D}^2\Bigg[1-\frac{\ln \left[ \ln
\left(1/n_{2D}
a_{2D}^{2}\right)\right]}{\ln\left(1/n_{2D}a_{2D}^{2}\right)}+\frac{B}{\ln\left(1/n_{2D}a_{2D}^{2}\right)}\nonumber  \\
&+&\left(\gamma_{2D} +2R_{2D}\frac{\text{arccoth}\left(
\sqrt{1+\frac{2t}{n_{2D}g_{2D}}} \right)}{\ln
(1/n_{2D}a_{2D}^{2})}\right)\Bigg],
\end{eqnarray}
where $L^2$ is the surface area of gas, $n_{2D}=n_0d$ is the surface
density and $B=1/2-\ln \left(mt/n_{2D}2\pi \hbar ^{2}\right)$. In
addition, the two parameters of disorder respectively take their 2D
value $\gamma_{2D}$ and $R_{2D}$. Both parameters, however, depend
the 2D expression of $\tilde{g}_{imp}$ which needs to be obtained
from investigating in detail the 2D scattering problem of a boson
with a quenched impurity. Such problem is definitely non-trivial,
and shall be left for further investigation in the future. In spite
of this, Eq. ({\ref{eg2d}}) has shed light on the ground state
properties of a 2D Bose gas in the presence of weak disorder.

Particularly, Eq. ({\ref{eg2d}}) presents one of the key results of
this paper as follows: First, Eq. ({\ref{eg2d}}) in the absence of
disorder formally reproduces corresponding results in Ref.
\cite{pilati} for the ground state energy of a purely 2D dilute Bose
gas. From this viewpoint, we expect that the character of a
1D-lattice-confined Bose gas in the presence of weak disorder in the
2D regime will be similar to a purely 2D Bose gas in a random
potential. Therefore, we argue that Eq. ({\ref{eg2d}}) provides an
analytical expression for the ground state energy of a uniform 2D
Bose gas in the presence of weak disorder. Specifically, the last
two terms provide the contribution of disorder to the ground state
energy. Second, Eq. (\ref{eg2d}) has provided beyond mean-field
corrections due to quantum fluctuations in the 2D geometry. These
corrections arise from combined effects of interatomic interaction
and disorder, and exhibit in 2D a characteristic
$1/\ln\left(1/n_{2D}a_{2D}^{2}\right)$ dependence, in contrast to
the 3D counterpart $\sqrt{n_0a_{3D}^{3}}$.

In a similar fashion, we analyze the asymptotic behavior of quantum
depletion. In the limit $2t/\tilde{g}_e n_0\gg 1$, corresponding to
the anisotropic 3D regime, $H(x)\simeq {2}/{(3\sqrt{x})}$.
Consequently, one finds the quantum depletion in 3D,
\begin{equation}  \label{DN3}
\frac{\Delta N}{N}\Bigg|_{3D}\simeq \left(\frac{8}{3}+\frac{\pi
}{2}\tilde{R}_{3D}\right)\sqrt{\frac{m^{*}}{m}}\left(\frac{n_{0}\tilde{a}_{3D}^{3}}{\pi
}\right)^{1/2},
\end{equation}
characterized by the dependence on the 3D gas parameter
$(n_{0}\tilde{a}_{3D}^{3})^{1/2}$. In the opposite 2D limit, on the
other hand, $\tilde{g}_e=g_{2D}d$ and $H(x)$ saturates to the value
$\pi /2$. Eq. (\ref{Depletion}) thereby asymptotically approaches
the 2D quantum depletion as
\begin{equation}  \label{DN2}
\frac{\Delta N}{N}\Bigg|_{2D}\simeq
\left(1+\frac{R_{2D}}{4}\right)\frac{1}{\ln
\left(1/n_{2D}a_{2D}^{2}\right)},
\end{equation}
which is proportional to $1/\ln \left(1/n_{2D}a_{2D}^{2}\right)$,
the small parameter in 2D. For varnishing disorder, Eq. (\ref{DN2})
is in good agreement with Ref. \cite{Schick,pilati} on the quantum
depletion of a purely weakly interacting 2D Bose gas. The second
term in Eq. ({\ref{DN2}}), therefore, presents the disorder-induced
condensate depletion in 2D. Furthermore, comparison of Eq.
(\ref{DN2}) with Eq. ({\ref{DN3}}) shows that, in the region where
Bogoliubov theory applies, for the same value of the gas parameter
the quantum depletion due to disorder is larger in 2D than in 3D.
Similar conclusion has been drawn in Ref. \cite{pilati} for the
quantum depletion induced by interatomic interaction.

\section{Superfluid Density}

In this section, we calculate the superfluid density of a dilute
Bose gas in the presence of a 1D optical lattice and weak disorder.
The general definition of the superfluid density is proposed by
Hohenberg and Martin \cite{Hohenberg}. We emphasize that
superfluidity is a kinetic property of a system and superfluid
density is essentially a transport coefficient, in contrast to the
condensate density which is an equilibrium quantity. Superfluid
density can be determined by the response of the system to an
external perturbation \cite{Hohenberg}.

In this paper, we adopt following definition: supposing that a linear phase
$\mathbf{Q}\cdot \mathbf{r}$ is imposed on the originally static
bosonic field which gives rise to a superfluid velocity $\mathbf{
\upsilon}=\hbar \mathbf{Q}/m$; in response, the thermodynamic
potential of the system is changed by \cite{Rey,Paan,legget}
\begin{equation}
\frac{\delta \Omega}{V}=\frac{\hbar ^{2}}{2m}\sum_{\alpha \beta}n
_{\alpha \beta }Q_{\alpha }Q_{\beta }. \label{Rhos}
\end{equation}
Here, the transport coefficient $n_{\alpha\beta}$ is interpreted as
the superfluid density \cite{legget}. In general, the
$n_{\alpha\beta}$ is a tensor for an anisotropic system.

To obtain $n_{\alpha\beta}$, we substitute the wavefunction for a
flowing condensate $\psi\left({\bf r},\tau\right)=\varphi\left({\bf r},\tau
\right)e^{i\mathbf{Q}\cdot {\bf r}}$ into Eq. ( \ref{Action}) and
obtain the action $S_{{\bf Q}}$ for the superfluid
\begin{equation}  \label{SQ}
S_{\mathbf{Q}}=S +{\hbar \beta
V}\sum\limits_{\mathbf{k},n}\psi^{*}_{\mathbf{k},n}
\left[f_{\mathbf{kQ}}+\frac{\hbar ^{2}}{2m}Q^{2}\right] \psi
_{\mathbf{k},n},
\end{equation}
where $S$ refers to the action for a static BEC in Eq.
({\ref{Action}}), and $f_{\mathbf{kQ}}=\left[\hbar
^{2}/m\left(k_{x}Q_{x}+k_{y}Q_{y}\right)+2Q_{z}td\sin
\left(k_{z}d\right)\right]$. Proceeding in a similar fashion as in
Sec. II, we obtain
\begin{eqnarray}
\Omega_{\mathbf{Q}} &=&V(-\tilde{\mu}n_{0}+n_{imp}\tilde{g}_{imp}n_{0}+\frac{%
\tilde{g}_{e}n_{0}^{2}}{2})  \notag \\
&&-\frac{1}{2}\sum\limits_{\mathbf{k\neq 0}}(\varepsilon _{\mathbf{k}}^{0}-%
\tilde{\mu}+2\tilde{g_{e}}n_{0}-\widetilde{E}_{\mathbf{k}})  \notag \\
&&-n_{imp}\tilde{g}_{imp}^{2}n_{0}\sum\limits_{\mathbf{k\neq 0}}\frac{%
\varepsilon _{\mathbf{k}}-\tilde{\mu}+\tilde{g}_{e}n_{0}}{\widetilde{E}_{\mathbf{k}%
}^{2}-f_{\mathbf{kQ}}^{2}}  \label{Deltaomega}
\end{eqnarray}
where $\widetilde{E}_{\mathbf{k}}=\sqrt{ (\varepsilon
_{\mathbf{k}}^0-\widetilde{\mu}
+2\tilde{g}_{e}n_{0})^{2}-\tilde{g}_{e} ^{2}n_{0}^{2}}$ depends on
${\bf Q}$ though $\tilde{\mu}=\mu-\hbar^{2}{\mathbf{Q}}^{2}/2m$.

Since the presence of a 1D optical lattice breaks the global
rotational symmetry and leaves the gas system only isotropic in the
$x-y$ plane, one can write
$n_{\alpha\beta}=n_{\alpha\alpha}\delta_{\alpha\beta}$ where
$n_{xx}=n_{yy}\neq n_{zz}$. Expanding Eq. (\ref{Deltaomega}) in
powers of ${\bf Q}$ and truncating at the quadratic order, we
compare the resulting expression with Eq. (\ref{Rhos}) and obtain
\begin{eqnarray}
n_{xx} &=&n_{yy}=n-\frac{2n_{imp}\tilde{g}_{imp}^{2}n_0}{V}
\sum\limits_{\mathbf{k\neq 0}}\frac{\hbar ^{2}k_{x}^{2}}{m}\frac{
\varepsilon^{0}_{\mathbf{k}}}{E_{\mathbf{k}}^4}, \label{Rhosx}
\end{eqnarray}
and
\begin{eqnarray}
n_{zz} &=&n\!-\!\frac{2mn_{imp}\tilde{g}_{imp}^{2}n_0}{\hbar ^{2}V}
\sum\limits_{\mathbf{k\neq
0}}\frac{\varepsilon^{0}_{\mathbf{k}}}{E_{\mathbf{
k}}^{4}}\left[2td\sin\left(k_{z}d\right)\right]^{2}.  \label{Rhosz}
\end{eqnarray}
Similar results have been obtained in Ref. \cite{Hu} using
current-current response function. The formal agreement between the
two affirms that, in spite of different ways to impose perturbation
and various options of physical quantities to measure the response,
these different routes to obtain superfluid density can be unified
within the framework of the linear response theory.

The disorder-induced normal fluid density fraction can be obtained
through
$\left(n_{n}\right)_{\alpha\beta}=\left(1-n_{\alpha\alpha}/n\right)\delta_{\alpha\beta}$.
Taking the continuum limit of Eqs. (\ref{Rhosx}) and
({\ref{Rhosz}}), one finds
\begin{eqnarray}
(n_n)_{xx}&=&(n_n)_{yy}=\tilde{R}\frac{m\tilde{g}_e}{8\hbar ^{2}\pi
d } I\left(\frac{2t}{\tilde{g}_en_0}\right),  \label{rouxx}
\end{eqnarray}
and
\begin{eqnarray}
(n_n)_{zz}&=&\tilde{R}\left(\frac{m}{
m^{*}}\right)^{2}\frac{1}{16\pi
n_0d^{3}}K\left(\frac{2t}{\tilde{g}_en_0}\right), \label{rouzz}
\end{eqnarray}
where $I(x)$ and $K(x)$ are functions of variable
$x=2t/\tilde{g}_en_{0}$ respectively defined as
\begin{equation}  \label{I}
I\left(x\right)=\left[\sqrt{1+x}-x\ln\left(\frac{1+\sqrt{1+x}}{\sqrt{x}}\ \right)\right],
\end{equation}
and
\begin{eqnarray}  \label{K}
K\left(x\right)=\ln\Bigg(\frac{1+\sqrt{1+x}}{\sqrt{x}}\Bigg)
-\frac{2-\left(2-x\right)\sqrt{1+x}}{x^{2}}.
\end{eqnarray}

\begin{figure}
\includegraphics[width=\columnwidth,height=50mm]{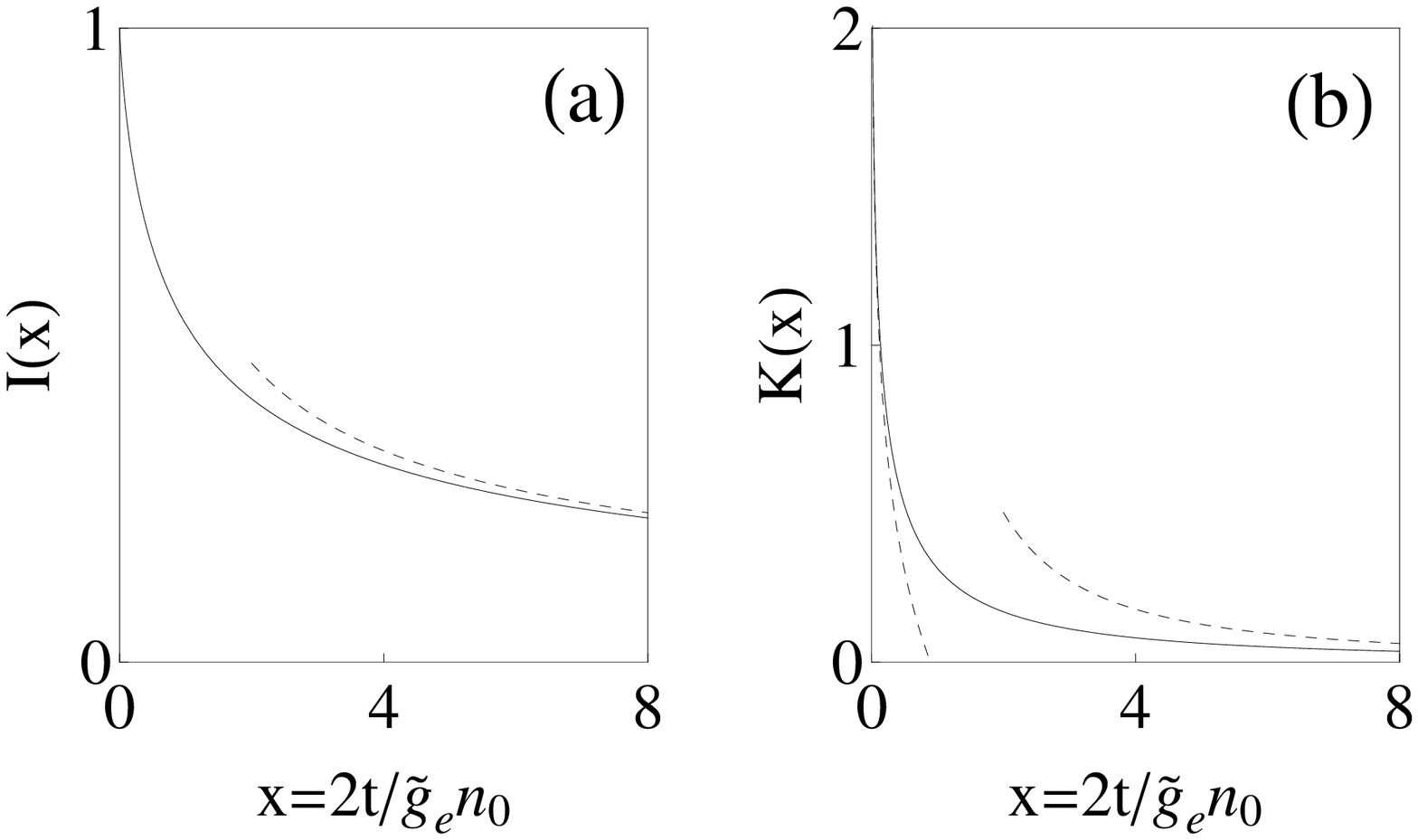}
\caption{\label{ik} (a): Scaling function I(x) in Eq. (\ref{I}) (solid line) and its
asymptotic behavior (dashed line). (b): Scaling function K(x) in Eq. (\ref{K}) (solid
line) and its asymptotic behavior (dashed line).}\label{ikk}
\end{figure}

The results of Eqs. (\ref{I}) and (\ref{K}) are plotted in Fig.
\ref{ikk}. In the asymptotic 3D limit, one finds $I(x)\simeq
2/3\sqrt{x}$ and $ K(x)\simeq 4 /3x^{3/2}$, corresponding the dashed
curves in Fig. \ref{ikk}. In such situation, Eqs. (\ref{rouxx}) and
( \ref{rouzz}) respectively become
\begin{eqnarray}
(n_n)_{xx}=(n_n)_{yy}&\simeq &\frac{2\pi
}{3}\tilde{R}_{3D}\sqrt{\frac{m^{\ast }}{m }}\left(
\frac{n_0\tilde{a}_{3D}^{3}}{\pi
}\right)^{\frac{1}{2}},\label{Rhoxx3D}
\end{eqnarray}
and
\begin{eqnarray}
(n_n)_{zz} &\simeq &\frac{2\pi
}{3}\tilde{R}_{3D}\sqrt{\frac{m}{m^{*}}}\left(\frac{
n_0\tilde{a}_{3D}^{3}}{\pi }\right)^{\frac{1}{2}}.  \label{Rhon3D}
\end{eqnarray}
Eq. ({\ref{Rhon3D}}) demonstrates similar dependence on the 3D gas
parameter as the 3D quantum depletion in Eq. ({\ref{DN3}}).
Moreover, the ratio between Eq. ({\ref{Rhoxx3D}}) and the disorder
induced quantum depletion $n_d$ in Eq. (\ref{DN3}) equals $4/3$ in
the unconfined X(Y)direction, in agreement with Ref. \cite{Huang};
whereas, this ratio becomes $(n_n)_{zz}/n_d=4m^{*}/3m$ due to the
increased inertia of the gas along the direction of optical lattice
\cite{Hu}.

In the opposite 2D limit, one obtains the limiting expression
$I(x)\simeq 1$ and $(n_n)_{2D}=(n_n)_{xx}=(n_n)_{yy}$ is found to be
\begin{equation}  \label{Nn2D}
(n_n)_{2D}\simeq \frac{R_{2D}}{2}\frac{1}{\ln\left(1/n_{2D}
a_{2D}^{2}\right)}.
\end{equation}
Equation ({\ref{Nn2D}}) presents another key result of this paper,
providing an analytical expression for the normal fluid density in a
homogenous Bose fluid in 2D in the presence of weak disorder. Eq.
({\ref{Nn2D}}) shows that the normal fluid density in 2D exhibits a
characteristic $1/\ln\left(1/n_{2D}a_{2D}^{2}\right)$ dependence.
With respect to the 3D case, a comparison of Eqs. ({\ref{DN2}}) and
(\ref{Nn2D}) leads to $n_n/n_d=2$ in 2D, indicating a more
pronounced contrast between superfluidity and Bose-Einstein
condensation at $T=0$. On the other hand, $K(x)$ in Eq. ({\ref{K}})
diverges in the limit $x\rightarrow 0$, leading to diverging
$n_{zz}$ Eq. ({\ref{rouzz}}) for vanishing tunneling. This signals
the absence of superfluidity along the direction of optical lattice,
which is consistent with the kinematical 2D nature of the Bose gas
in the absence of tunneling along the direction of the laser.

\section{Possible experimental scenarios and conclusion}

Central to testing the validity of the physics in this article
concerns experimental realization of a BEC in the superfluid phase
along the entire evolution from 3D to quasi-2D. Present facilities
have allowed one to adjust the depth of an lattice, realize tight
confinement of the motion of trapped particles and ultimately
achieve a kinematically 2D gas. In typical experiments to date,
quasi-2D quantum degenerate Bose gases have been experimentally
produced both in single ¡®pancake¡¯ traps and at the nodes of 1D
optical lattice potentials \cite{Exp2D}. In addition, it has been
suggested that BEC and superfluidity can be both achieved below a
critical temperature \cite{Petrov}. Furthermore, adding a tunable
periodic potential allows one to combine the benefit of the reduced
dimensionality with the advantage of working with large yet coherent
samples \cite{Orso}.

Upon overcoming above difficulties, the experimental realization of
our scenario amounts to controlling three parameters whose interplay
underlies the physics of this work: the strength of an optical
lattice $s$, the interaction between bosonic atoms $\tilde{g}n_0$,
and the strength of disorder $\tilde{R}$. All these quantities are
experimentally controllable using state-of-the-art technologies. The
interatomic interaction can be controlled in a very versatile manner
via the technology of Feshbach resonances \cite{Fesh}. In the typical
experiments to date, the values of ratio $\tilde{g}n_0/E_R$ range
from $0.02$ to $1$ \cite{Morschrev,Blochrev}. The depth of an
optical lattice $s$ can be changed from $0E_R$ to $32E_R$ almost at
will \cite{Greiner}. Disorder may be created in a repeatable way by
introducing impurities in the sample \cite{Ospelkaus}, or using
laser speckles and multi-chromatic lattices
\cite{Roth,Bamski,Gavish}.

Further difficulties may arise in measuring the beyond-mean-field
corrections to the ground state energy along the dimensional
crossover. For typical values of the atom density and scattering
length, such corrections remain very small and hard to observe in
usual experiments that measure density profiles or release energy.
They can be visible, however, in the frequencies of collective
excitations in a lattice system \cite{ Orso,Pita,Altmeyer}. The
direct measurement of quantum depletions of a quasi-2D condensate
can be achieved either through observing ballistic expansion
\cite{Xu} or applying Bragg spectroscopy \cite{Bragg}. It is worth
mentioning that the possibility to use ballistic expansion to
measure quantum fluctuations is associated with the characteristics
of an optical lattice where the confinement frequency at each
lattice site far exceeds the interaction energy. As such, the
time-of-flight images are essentially a snapshot of the frozen-in
momentum distribution of the wavefunction at the time of the lattice
switch-off, thus allowing for a direct observation of quantum
depletions. This technology cannot be applied, for example, to
measure quantum depletions of a quasi-2D Bose gas confined in a
harmonic trap. From this perspective, Bragg spectroscopy admits
broader ranges of application, independent of methods of confinement
to create quasi-2D BEC's systems.

We expect, therefore, that the phenomena discussed in this article
should be observable within current experimental capability. We
emphasize here that the presented work is restricted to weak
disorder and weak interatomic interaction. For further
investigations in the presence of stronger interatomic interaction
or disorder, the path-integral Monte Carlo simulation is a reliable
method \cite{MC}.

In summary, we have investigated a dilute Bose gas trapped in a 1D
optical lattice and a random potential. Capitalizing on the
characteristic dimensional crossover properties, the obtained
results in the quasi-2D regime allow us to derive analytical
expressions for the ground state energy, quantum depletion and
superfluid density of an effectively pure 2D Bose gas in the
presence of weak disorder. Our analysis signifies a more pronounced
effect of disorder in systems with reduced dimensionality in
enhancing quantum fluctuations and depleting superfluid density. In
particular, the ratio between the normal fluid density and the
corresponding condensate depletion increases to 2 in 2D, in contrast
to the familiar $4/3$ in lattice-free 3D geometry.

\bigskip
\textbf{Acknowledgements}
This work is supported by the NSF of China (Grant No. 10674139).
H.Y. is supported by the Hongkong Research Council (RGC) and
the Hong Kong Baptist University Faculty Research Grant (FRG).
L.Z.X. is supported by the IMR SYNL-TS K\^e Research Grant.

\end{document}